# New Binomial Bent Function over the Finite Fields of Odd Characteristic*

Tor Helleseth and Alexander Kholosha

The Selmer Center
Department of Informatics, University of Bergen
P.O. Box 7800, N-5020 Bergen, Norway
{Tor.Helleseth,Alexander.Kholosha}@uib.no

**Abstract.** The $p$-ary function $f(x)$ mapping $\mathrm{GF}(p^{4k})$ to $\mathrm{GF}(p)$ given by $f(x) = \mathrm{Tr}_{4k}\big(x^{p^{3k}+p^{2k}-p^k+1} + x^2\big)$ is proven to be a weakly regular bent function and the exact values of its Walsh transform coefficients are found. The proof is based on a few new results in the area of exponential sums and polynomials over finite fields that may also be interesting as independent problems.

## 1 Introduction

Boolean bent functions were first introduced by Rothaus in 1976 as an interesting combinatorial object with the important property of having the maximum Hamming distance to the set of all affine functions. Later the research in this area was stimulated by the significant relation to the following topics in computer science: coding theory, sequences and cryptography (design of stream ciphers and $S$-boxes for block ciphers). Kumar, Scholtz and Welch in [1] generalized the notion of Boolean bent functions to the case of functions over an arbitrary finite field. Complete classification of bent functions looks hopeless even in the binary case. In the case of generalized bent functions things are naturally much more complicated. However, many explicit methods are known for constructing bent functions either from scratch or based on other, simpler bent functions.

Given a function $f(x)$ mapping $\mathrm{GF}(p^n)$ to $\mathrm{GF}(p)$, the direct and inverse *Walsh transform* operations on $f$ are defined at a point by the following respective identities:

$$S_f(b) = \sum_{x\in\mathrm{GF}(p^n)} \omega^{f(x)-\mathrm{Tr}_n(bx)} \quad \text{and} \quad \omega^{f(x)} = \frac{1}{p^n}\sum_{b\in\mathrm{GF}(p^n)} S_f(b)\omega^{\mathrm{Tr}_n(bx)}$$

where $\mathrm{Tr}_n() : \mathrm{GF}(p^n) \to \mathrm{GF}(p)$ denotes the absolute trace function, $\omega = e^{\frac{2\pi i}{p}}$ is the complex primitive $p^{\mathrm{th}}$ root of unity and elements of $\mathrm{GF}(p)$ are considered as integers modulo $p$.

* This work was supported by the Norwegian Research Council and partially by the grant NIL-I-004 from Iceland, Liechtenstein and Norway through the EEA and Norwegian Financial Mechanisms.

According to [1], $f(x)$ is called a *p-ary bent function* (or *generalized bent function*) if all its Walsh coefficients satisfy $|S_f(b)|^2 = p^n$. A bent function $f(x)$ is called *regular* (see [1, Definition 3] and [2, p. 576]) if for every $b \in \mathrm{GF}(p^n)$ the normalized Walsh coefficient $p^{-n/2}S_f(b)$ is equal to a complex $p^{\text{th}}$ root of unity, i.e., $p^{-n/2}S_f(b) = \omega^{f^*(b)}$ for some function $f^*$ mapping $\mathrm{GF}(p^n)$ into $\mathrm{GF}(p)$. A bent function $f(x)$ is called *weakly regular* if there exists a complex $u$ having unit magnitude such that $up^{-n/2}S_f(b) = \omega^{f^*(b)}$ for all $b \in \mathrm{GF}(p^n)$. We call $u^{-1}p^{n/2}$ the *magnitude* of $S_f(b)$. Throughout this paper, $p^{n/2}$ with odd $n$ stands for the *positive* square root of $p^n$. For a comprehensive reference on monomial and quadratic $p$-ary bent functions we refer reader to [3].

In the present paper, we take an odd prime $p$ and examine prospective $p$-ary bent functions having the form $f(x) = \mathrm{Tr}_n\left(a_0x^{d_0} + a_1x^{d_1}\right)$ with $a_i, x \in \mathrm{GF}(p^n)$ and arbitrary integer exponents $d_i$ for $i = \{0, 1\}$. Functions of this type with both coefficients $a_i$ being nonzero are called *binomial.* We prove that $f(x) = \mathrm{Tr}_n\left(x^{p^{3k}+p^{2k}-p^k+1} + x^2\right)$ is a weakly regular bent function for $n = 4k$ and find the exact values of its Walsh transform coefficients. It is interesting that in the binary case when $p = 2$, the decimation $2^{3k} - 2^{2k} + 2^k + 1$ which is cyclotomic equivalent to the exponent in the first term of the above bent function, was studied by Niho in [4, Theorem ???] and Helleseth in [5]. They proved that the cross-correlation function between two binary $m$-sequences given by this decimation is four-valued and found the distribution.

Note that the only not weakly regular bent function known so far is the power function from [3, Fact 1]. Here we present a binomial with this property.

**Fact 1** *The ternary function $f(x)$ mapping $\mathrm{GF}(3^4)$ to $\mathrm{GF}(3)$ given by $f(x) = \mathrm{Tr}_4\left(a_0x^{22} + x^4\right)$, where $a_0 \in \{\xi^{10}, \xi^{30}\}$ and $\xi$ is a primitive element of $\mathrm{GF}(3^4)$, is bent and not weakly regular bent.*

It is interesting to note that $22 = 3^{3k} - 3^{2k} + 3^k + 1$ for $k = 1$. Unfortunately, we were not able to find a family containing the function from Fact 1. Naturally, more binomial bent functions of these types can be constructed using cyclotomic equivalence of exponents and coefficients.

## 2 Preliminary Results

In this section, we present some preliminary results. However, they may also be interesting as independent finite field problems. The finite field $\mathrm{GF}(p^k)$ is a subfield of $\mathrm{GF}(p^n)$ if and only if $k$ divides $n$. The trace mapping from $\mathrm{GF}(p^n)$ to the subfield $\mathrm{GF}(p^k)$ is defined by $\mathrm{Tr}_k^n(x) = \sum_{i=0}^{n/k-1} x^{p^{ik}}$. In the case when $k = 1$, we use the notation $\mathrm{Tr}_n(x)$ instead of $\mathrm{Tr}_1^n(x)$.

**Proposition 1.** *Let $n = 4k$ and for any $a \in \mathrm{GF}(p^n)$ define*

$$C(a) = \sum_{x \in \mathrm{GF}(p^n)} \omega^{\mathrm{Tr}_n\left(x + ax^{(p^{2k}+1)/2}\right)} \quad . \tag{1}$$

*Then*

$$C(a)-C(-a)=\begin{cases} sp^{3k}, & \text{if } a+a^{p^{2k}}=-2\\ -sp^{3k}, & \text{if } a+a^{p^{2k}}=2\\ 0, & \text{otherwise}\end{cases},$$

*where* $s=(-1)^k$ *if* $p\equiv 3 \pmod 4$ *and* $s=1$ *otherwise.*

*Proof.* Denote $d=(p^{2k}+1)/2$ which is odd. First, note that $d=(p^{2k}-1)/2+1\equiv 1 \pmod{p-1}$ and, therefore, $tx+a(tx)^d=t(x+ax^d)$ for any $t\in\mathrm{GF}(p)$. This means that all nonzero elements in $\{\mathrm{Tr}_n(x+ax^d)\mid x\in\mathrm{GF}(p^n)\}$ appear equally often and $C(a)$ is an integer.

We can calculate

$$\begin{aligned}\sum_{a\in\mathrm{GF}(p^n)} C(a)^2 &= \sum_{a,x,y\in\mathrm{GF}(p^n)} \omega^{\mathrm{Tr}_n\left(x+y+a(x^d+y^d)\right)}\\ &= \sum_{x,y\in\mathrm{GF}(p^n)} \omega^{\mathrm{Tr}_n\left(x+y\right)} \sum_{a\in\mathrm{GF}(p^n)} \omega^{\mathrm{Tr}_n\left(a(x^d+y^d)\right)}\\ &\overset{(*)}{=} p^n+p^n \sum_{z\in\mathrm{GF}(p^n),\, z^d=1}\ \sum_{x\in\mathrm{GF}(p^n)^*} \omega^{\mathrm{Tr}_n\left(x(1-z)\right)}\\ &= p^n+p^n(p^n-1)-p^n\#\{z\in\mathrm{GF}(p^n)\mid z^d=1,\ z\neq 1\}\\ &= p^{2n}+p^n(1-d)\ ,\end{aligned}$$

where in $(*)$ we made a substitution $y=-zx$. Similarly,

$$\begin{aligned}\sum_{a\in\mathrm{GF}(p^n)} C(a)C(-a) &= p^n+p^n \sum_{z\in\mathrm{GF}(p^n),\, z^d=1}\ \sum_{x\in\mathrm{GF}(p^n)^*} \omega^{\mathrm{Tr}_n\left(x(1+z)\right)}\\ &= p^n-p^n\#\{z\in\mathrm{GF}(p^n)\mid z^d=1\}=p^n(1-d)\ ,\end{aligned}$$

where $y=zx$. Therefore,

$$\sum_{a\in\mathrm{GF}(p^n)}(C(a)-C(-a))^2=2\left(\sum_{a\in\mathrm{GF}(p^n)} C(a)^2-\sum_{a\in\mathrm{GF}(p^n)} C(a)C(-a)\right)=2p^{2n}\ .$$

Further, since any $a\in\mathrm{GF}(p^n)$ with $\mathrm{Tr}^n_{2k}(a)=a+a^{p^{2k}}=2$ can be uniquely written as $a=1+b$ for some $b\in\mathrm{GF}(p^n)$ with $\mathrm{Tr}^n_{2k}(b)=0$, we can write

$$\sum_{a\in\mathrm{GF}(p^n),\,\mathrm{Tr}^n_{2k}(a)=2} C(a)=\sum_{x\in\mathrm{GF}(p^n)} \omega^{\mathrm{Tr}_n\left(x+x^d\right)} \sum_{b\in\mathrm{GF}(p^n),\,\mathrm{Tr}^n_{2k}(b)=0} \omega^{\mathrm{Tr}_n\left(bx^d\right)}. \quad (2)$$

If $x$ is a nonsquare then select $r\in\mathrm{GF}(p^n)^*$ with $r^{p^{2k}-1}=-1$ (note that such $r$ is a nonsquare in $\mathrm{GF}(p^n)$ since $d$ is odd) and write $x=ry^2$ for $y\in\mathrm{GF}(p^n)^*$. Then

$$\begin{aligned}\mathrm{Tr}_n\left(bx^d\right) &= \mathrm{Tr}_n\left(br^d y^{p^{2k}+1}\right)=\mathrm{Tr}_{2k}\left(y^{p^{2k}+1}(br^d+b^{p^{2k}}r^{dp^{2k}})\right)\\ &= \mathrm{Tr}_{2k}\left(y^{p^{2k}+1}(br^d+(-b)(-r)^d)\right)=2\mathrm{Tr}_{2k}\left(br^d y^{p^{2k}+1}\right)\end{aligned}$$

Since $br^d$ runs through all $\mathrm{GF}(p^{2k})$ when $b$ runs through the elements of $\mathrm{GF}(p^n)$ with $\mathrm{Tr}^n_{2k}(b) = 0$, then the inner sum in (2) is equal to zero. If $x = y^2$ is a square in $\mathrm{GF}(p^n)$ then $x^d \in \mathrm{GF}(p^{2k})$ and $\mathrm{Tr}_n\big(bx^d\big) = \mathrm{Tr}_{2k}\big(x^d \mathrm{Tr}^n_{2k}(b)\big) = 0$. Therefore the sum in (2) is equal to

$$
\begin{aligned}
&\frac{p^{2k}}{2} \sum_{y \in \mathrm{GF}(p^n)} \omega^{\mathrm{Tr}_n\left(y^2 + y^{p^{2k}+1}\right)} \\
&\overset{(*)}{=} \frac{p^{2k}}{2} + \frac{p^{2k}}{2} \sum_{r \in R} \sum_{z \in \mathrm{GF}(p^{2k})^*} \omega^{\mathrm{Tr}_{2k}\left(z^2 \mathrm{Tr}^n_{2k}\left(r^2 + r^{p^{2k}+1}\right)\right)} \\
&= \frac{p^{2k}}{2} + \frac{p^{2k}}{2} \sum_{r \in R} \sum_{z \in \mathrm{GF}(p^{2k})^*} \omega^{\mathrm{Tr}_{2k}\left(z^2 \left(r + r^{p^{2k}}\right)^2\right)} \\
&= \frac{p^{2k}}{2} - \frac{p^{2k}}{2} \sum_{r \in R,\, r \neq \xi^d} \Big(\eta\big((r + r^{p^{2k}})^2\big) s p^k + 1\Big) + \frac{p^{2k}(p^{2k}-1)}{2} \\
&= -\frac{sp^{5k}}{2} \ ,
\end{aligned}
$$

where in $(*)$ we made a substitution $y = rz$ with $r \in R = \{\xi^i \mid i = 0, \ldots, p^{2k}\}$ and $z \in \mathrm{GF}(p^{2k})^*$, $\xi$ being a primitive element of $\mathrm{GF}(p^n)$. At the end, we used [3, Corollary 3] assuming $s = (-1)^k$ if $p \equiv 3 \pmod 4$ and $s = 1$ otherwise, $\eta(\cdot)$ is the quadratic character of $\mathrm{GF}(p^{2k})$. Also note that $r + r^{p^{2k}} = 0$ only for $r = \xi^d$.

Similarly,

$$
\sum_{a \in \mathrm{GF}(p^n),\, \mathrm{Tr}^n_{2k}(a) = 2} C(-a) = -\frac{p^{2k}}{2} \sum_{r \in R,\, r \neq 1} \Big(\eta\big((r - r^{p^{2k}})^2\big) s p^k + 1\Big) + \frac{p^n}{2} = \frac{sp^{5k}}{2}
$$

since $r - r^{p^{2k}} = 0$ only for $r = 1$. Also observe that $(r - r^{p^{2k}})^2$ is a nonsquare in $\mathrm{GF}(p^{2k})$ for any $r \in R$ except when $r = 1$, which can be seen from the following. If $t = r - r^{p^{2k}} \neq 0$ then $t^{p^{2k}} = -t$ and $t = \xi^i$ for some $i \equiv d \pmod{p^{2k}+1}$. Thus, $t^2 = \xi^{(p^{2k}+1)(2j+1)} = \gamma^{2j+1}$, where $\gamma$ is a primitive element of $\mathrm{GF}(p^{2k})$.

From the identities proven earlier it follows that

$$
\sum_{a \in \mathrm{GF}(p^n),\, \mathrm{Tr}^n_{2k}(a) = 2} (C(a) - C(-a)) = \sum_{a \in \mathrm{GF}(p^n),\, \mathrm{Tr}^n_{2k}(a) = -2} (C(-a) - C(a)) = -sp^{5k}.
$$

Since $C(a)$ is an integer, the average value of $C(a) - C(-a)$ over all the elements $a \in \mathrm{GF}(p^n)$ with $\mathrm{Tr}^n_{2k}(a) = 2$ (resp. $\mathrm{Tr}^n_{2k}(a) = -2$) is equal to $-sp^{3k}$ (resp. $sp^{3k}$) and we obtain

$$
\sum_{a \in \mathrm{GF}(p^n),\, \mathrm{Tr}^n_{2k}(a) = \pm 2} (C(a) - C(-a))^2 \geq 2p^{2n} = \sum_{a \in \mathrm{GF}(p^n)} (C(a) - C(-a))^2 \ .
$$

Thus, the above inequality becomes an identity leading to $C(a) = C(-a)$ for any $a \in \mathrm{GF}(p^n)$ with $\mathrm{Tr}^n_{2k}(a) \neq \pm 2$. Also,

$$
\sum_{a \in \mathrm{GF}(p^n),\, \mathrm{Tr}^n_{2k}(a) = 2} (C(a) - C(-a))^2 = \sum_{a \in \mathrm{GF}(p^n),\, \mathrm{Tr}^n_{2k}(a) = -2} (C(a) - C(-a))^2 = p^{2n}
$$

that is the lowest possible value of these sums of squares and which is possible only if $C(a) - C(-a) = -sp^{3k}$ (resp. $C(a) - C(-a) = sp^{3k}$) when $\mathrm{Tr}^n_{2k}(a) = 2$ (resp. $\mathrm{Tr}^n_{2k}(a) = -2$). $\square$

Let $n$ be even and $U$ denote a cyclic subgroup of order $p^{n/2}+1$ of the multiplicative group of $\mathrm{GF}(p^n)$ (generated by $\xi^{p^{n/2}-1}$, where $\xi$ is a primitive element of $\mathrm{GF}(p^n)$). Denote subsets of $U$ containing squares and non-squares respectively as

$$U_+ = \{u \in U \mid u^{(p^{n/2}+1)/2} = 1\} \quad \text{and} \quad U_- = \{u \in U \mid u^{(p^{n/2}+1)/2} = -1\} \ .$$

Also, if $n = 4k$ then for any $c \in \mathrm{GF}(p^k)$ and any fixed $b \in \mathrm{GF}(p^n)$ denote

$$n_+(c) = \#\{u \in U_+ \mid \mathrm{Tr}^n_k(bu) = c\} \quad \text{and} \quad n_-(c) = \#\{u \in U_- \mid \mathrm{Tr}^n_k(bu) = c\} \ .$$

Note that $(-1)^{(p^{2k}+1)/2} = -1$ and thus, $-U_+ = U_-$.

**Proposition 2.** *Let $n = 4k$ and take any $b \in \mathrm{GF}(p^n)$. Then $n_+(c) = n_-(c)$ for any $c \in \mathrm{GF}(p^k)$ except when $\mathrm{Tr}^n_k\big(b^{(p^{2k}+1)/2}\big) \neq 0$ and $c = \pm\mathrm{Tr}^n_k\big(b^{(p^{2k}+1)/2}\big)$. If $c = \mathrm{Tr}^n_k\big(b^{(p^{2k}+1)/2}\big) \neq 0$ then $n_+(c) - n_-(c) = sp^k$, where $s = (-1)^k$ if $p \equiv 3 \pmod 4$ and $s = 1$ otherwise.*

*Proof.* The claimed statement is obvious for $b = 0$ so further on assume $b \neq 0$. Denoting $d = (p^{2k}+1)/2$ we can write

$$\begin{aligned}
p^{5k} n_+(c) &= \sum_{u \in \mathrm{GF}(p^n)} \sum_{x \in \mathrm{GF}(p^n)} \omega^{\mathrm{Tr}_n\left(x(u^d-1)\right)} \sum_{y \in \mathrm{GF}(p^k)} \omega^{\mathrm{Tr}_k\left(y(\mathrm{Tr}^n_k(bu)-c)\right)} \\
&= \sum_{x \in \mathrm{GF}(p^n)} \sum_{y \in \mathrm{GF}(p^k)} \omega^{-\mathrm{Tr}_n\left(cy/4+x\right)} \sum_{u \in \mathrm{GF}(p^n)} \omega^{\mathrm{Tr}_n\left(ybu+xu^d\right)} \\
&\overset{(*)}{=} \sum_{z \in \mathrm{GF}(p^n)} \sum_{y \in \mathrm{GF}(p^k)^*} \omega^{-\mathrm{Tr}_n\left(y(c/4+z)\right)} \sum_{v \in \mathrm{GF}(p^n)} \omega^{\mathrm{Tr}_n\left(v+z(v/b)^d\right)} \\
&\quad + \sum_{x \in \mathrm{GF}(p^n)} \sum_{u \in \mathrm{GF}(p^n)} \omega^{\mathrm{Tr}_n\left(x(u^d-1)\right)} \\
&= \sum_{z \in \mathrm{GF}(p^n)} C\left(zb^{-d}\right) \sum_{y \in \mathrm{GF}(p^k)^*} \omega^{-\mathrm{Tr}_k\left(y(c+\mathrm{Tr}^n_k(z))\right)} + p^n d \\
&= p^k \sum_{\mathrm{Tr}^n_k(z)=-c} C\left(zb^{-d}\right) - \sum_{z \in \mathrm{GF}(p^n)} C\left(zb^{-d}\right) + p^n d \ ,
\end{aligned}$$

where in $(*)$ we made substitutions $v = ybu$ and $x = yz$ and using that $y^d = y$ for $y \in \mathrm{GF}(p^k)$; exponential sum $C(\cdot)$ comes from (1). Similarly,

$$p^{5k} n_-(c) = p^k \sum_{\mathrm{Tr}^n_k(z)=c} C\left(zb^{-d}\right) - \sum_{z \in \mathrm{GF}(p^n)} C\left(zb^{-d}\right) + p^n d$$

and, therefore, by Proposition 1,

$$p^n(n_+(c)-n_-(c)) = \sum_{\mathrm{Tr}_k^n(z)=-c} \left(C\left(zb^{-d}\right)-C\left(-zb^{-d}\right)\right)$$

$$= sp^{3k}\#\left\{z\in \mathrm{GF}(p^n)\mid \mathrm{Tr}_k^n(z)=-c \text{ and } \mathrm{Tr}_{2k}^n\left(zb^{-d}\right)=-2\right\}$$

$$-\ sp^{3k}\#\left\{z\in \mathrm{GF}(p^n)\mid \mathrm{Tr}_k^n(z)=-c \text{ and } \mathrm{Tr}_{2k}^n\left(zb^{-d}\right)=2\right\}$$

$$= sp^{3k}(m_-(-c)-m_+(-c))\ ,$$

where $m_+(c)$ and $m_-(c)$ are defined implicitly.

First, consider the case when $b$ is a square in $\mathrm{GF}(p^n)$ that holds if and only if $b^d\in \mathrm{GF}(p^{2k})$. Then,

$$\mathrm{Tr}_{2k}^n\left(zb^{-d}\right)=\left(z+z^{p^{2k}}\right)b^{-d}=\pm 2$$

is equivalent to $z+z^{p^{2k}}=\pm 2b^d$ giving

$$\mathrm{Tr}_k^n(z)=\pm 2\left(b^d+b^{p^kd}\right)=\pm\mathrm{Tr}_k^n\left(b^d\right)\ .$$

Note that $z+z^{p^{2k}}=a$ has $p^{2k}$ solutions in $\mathrm{GF}(p^n)$ for any $a\in\mathrm{GF}(p^{2k})$. If $c\neq\pm\mathrm{Tr}_k^n\left(b^d\right)$ then $m_+(-c)=m_-(-c)=0$. If $c=\mathrm{Tr}_k^n\left(b^d\right)=0$ then $m_+(-c)=m_-(-c)=p^{2k}$. If $-c=\mathrm{Tr}_k^n\left(b^d\right)\neq 0$ then $m_+(-c)=p^{2k}$ and $m_-(-c)=0$, if $-c=-\mathrm{Tr}_k^n\left(b^d\right)\neq 0$ then $m_+(-c)=0$ and $m_-(-c)=p^{2k}$.

Finally, if $b$ is a nonsquare in $\mathrm{GF}(p^n)$, i.e., $b^{(p^n-1)/2}=-1$, that holds if and only if $\mathrm{Tr}_{2k}^n\left(b^d\right)=0$, then

$$\mathrm{Tr}_{2k}^n\left(zb^{-d}\right)=\left(z-z^{p^{2k}}\right)b^{-d}=\pm 2$$

is equivalent to $z-z^{p^{2k}}=\pm 2b^d$. We can write

$$p^{5k}m_+(c) = \sum_{z\in\mathrm{GF}(p^n)}\sum_{x\in\mathrm{GF}(p^n)}\omega^{\mathrm{Tr}_n\left(x(z-z^{p^{2k}}-2b^d)\right)}\sum_{y\in\mathrm{GF}(p^k)}\omega^{\mathrm{Tr}_k\left(y(\mathrm{Tr}_k^n(z)-c)\right)}$$

$$= \sum_{x\in\mathrm{GF}(p^n)}\sum_{y\in\mathrm{GF}(p^k)}\omega^{-\mathrm{Tr}_n\left(cy/4+2xb^d\right)}\sum_{z\in\mathrm{GF}(p^n)}\omega^{\mathrm{Tr}_n\left(z(y+x-x^{p^{2k}})\right)}$$

$$\overset{(*)}{=} p^n\sum_{x\in\mathrm{GF}(p^n),\,L(x)=0}\omega^{-\mathrm{Tr}_n\left(c(x^{p^{2k}}-x)/4+2xb^d\right)}$$

$$= p^n\sum_{x\in\mathrm{GF}(p^n),\,L(x)=0}\omega^{-\mathrm{Tr}_n\left(2xb^d\right)}$$

$$= p^n\sum_{x\in\mathrm{GF}(p^n),\,L(x)=0}\omega^{\mathrm{Tr}_n\left(2xb^d\right)}$$

$$= p^{5k}m_-(c)\ ,$$

where in $(*)$ we used that $y=x^{p^{2k}}-x\in\mathrm{GF}(p^k)$ that is equivalent to $L(x)=x-x^{p^k}-x^{p^{2k}}+x^{p^{3k}}=0$. □

Note that $n_+(c) - n_-(c) = -sp^k$ if $c = -\mathrm{Tr}_k^n\left(b^{(p^{2k}+1)/2}\right) \neq 0$.

**Proposition 3.** *Let $n = 4k$ and take any $c \in \mathrm{GF}(p^k)^*$. The polynomial*

$$T_+(X,c) = X^{(p^{2k}+1)/2} + (X+c)^{(p^{2k}+1)/2} + X^{p^k(p^{2k}+1)/2} + (X+c)^{p^k(p^{2k}+1)/2}$$

*splits over $\mathrm{GF}(p^n)$ and for any root $x$ of $T_+(X,c)$, both $x$ and $x+c$ are squares in $\mathrm{GF}(p^n)$ and $x \neq 0$. On the other hand, the polynomial*

$$T_-(X,c) = X^{(p^{2k}+1)/2} - (X+c)^{(p^{2k}+1)/2} + X^{p^k(p^{2k}+1)/2} - (X+c)^{p^k(p^{2k}+1)/2}$$

*does not have any roots in the set of squares of $\mathrm{GF}(p^n)$. Moreover, for different values of $c$, the corresponding roots of $T_+(X,c)$ belong to disjoint subsets.*

*Proof.* First, we find the number of roots of $T_+(X,c)$ in $\mathrm{GF}(p^n)$. Note that all roots of $T_+(X,c)$ and $T_-(X,c)$ are exactly the roots of respectively $T_+(X,1)$ and $T_-(X,1)$ multiplied by $c$. Moreover, any $c \in \mathrm{GF}(p^k)^*$ is a square in $\mathrm{GF}(p^n)$. Therefore, it is sufficient to assume $c = 1$. Also note that

$$T_+(0,c) = c^{(p^{2k}+1)/2} + c^{p^k(p^{2k}+1)/2} = 2c \neq 0 \ .$$

Here we use the representation of $x \in \mathrm{GF}(p^n)$ suggested in [6, Theorem 11]. After fixing any $t \in \mathrm{GF}(p^n)^*$, we can write $x = \alpha + t\alpha^{-1}$, where $\alpha$ and $t\alpha^{-1}$ are the two roots in $\mathrm{GF}(p^{2n})^*$ of $z^2 - xz + t = 0$. Note that $x = \alpha_1 + t\alpha_1^{-1} = \alpha_2 + t\alpha_2^{-1}$ gives $(\alpha_1 - \alpha_2)\left((\alpha_1\alpha_2)^{-1}t - 1\right) = 0$ which holds if and only if $\alpha_1 = \alpha_2$ or $\alpha_1\alpha_2 = t$. Therefore, exactly two different values of $\alpha$ give the same $x$, unless $\alpha^2 = t$.

In our case, take $t = v^2$, where

$$v = \begin{cases} \frac{p-1}{4}, & \text{if } p \equiv 1 \pmod 4 \\ -\frac{p+1}{4}, & \text{if } p \equiv 3 \pmod 4 \ , \end{cases}$$

which means $2v = (p-1)/2$. Denote $d = (p^{2k}+1)/2$. Replacing $x$ with $x + (p-1)/2$ and using the above substitution for $x$ we obtain

$$\begin{aligned} T_+(x,1) &= \left(x + \frac{p-1}{2}\right)^d + \left(x - \frac{p-1}{2}\right)^d + \left(x + \frac{p-1}{2}\right)^{dp^k} + \left(x - \frac{p-1}{2}\right)^{dp^k} \\ &= \frac{(\alpha+v)^{2d} + (\alpha-v)^{2d}}{\alpha^d} + \left(\frac{(\alpha+v)^{2d} + (\alpha-v)^{2d}}{\alpha^d}\right)^{p^k} \\ &= 2\left(\alpha^d + v^2\alpha^{-d} + \alpha^{dp^k} + v^2\alpha^{-dp^k}\right) \\ &= 2\left(\alpha^d + \alpha^{dp^k}\right)\left(v^2\alpha^{-d(p^k+1)} + 1\right) \ , \end{aligned}$$

since $v^p = v$. Thus, $T_+(x,1) = 0$ if and only if

$$\alpha^{d(p^k-1)} = -1 \quad \text{or} \quad \alpha^{d(p^k+1)} = -v^2 \ . \qquad (3)$$

In particular, this means respectively that $\alpha^{p^n-1} = (-1)^{2(p^k+1)} = 1$ or $\alpha^{p^n-1} = (-v^2)^{2(p^k-1)} = 1$ and for this reason, $\alpha \in \mathrm{GF}(p^n)$. Therefore, $x = \alpha + v^2\alpha^{-1} \in \mathrm{GF}(p^n)$ for any $\alpha$ with $T_+(\alpha + v^2\alpha^{-1}, 1) = 0$.

Let $\xi$ be a primitive element of $\mathrm{GF}(p^n)$. Equation $\alpha^{d(p^k-1)} = -1 = \xi^{(p^n-1)/2}$ has $d(p^k-1)$ solutions and $\alpha^{d(p^k+1)} = -v^2 = \xi^{j(p^{2k}+1)(p^k+1)}$ (for some $j \in \{0, \cdots, p^k-2\}$) has $d(p^k+1)$ solutions in $\mathrm{GF}(p^n)$. The solution sets of these two equations are disjoint because if $\alpha^{d(p^k-1)} = -1$ then $\alpha^{d(p^k+1)} = -\alpha^{2d} \neq -v^2$ since otherwise, $\alpha^{d(p^k-1)} = v^{p^k-1} = 1$ giving a contradiction. In total, there are $2dp^k = p^k(p^{2k}+1)$ solutions of $T_+(\alpha + v^2\alpha^{-1}, 1) = 0$. Finally, we have to divide this number by 2 since exactly two values of $\alpha$ map to the same $x$ unless $\alpha^2 = v^2$ (in the latter case, $\alpha^{p^{2k}+1} = v^{p^{2k}+1} = v^2$ so $\alpha^{d(p^k-1)} = v^{p^k-1} = 1$ and $\alpha^{d(p^k+1)} = v^{p^k+1} = v^2$ giving $T_+(\alpha+v^2\alpha^{-1}, 1) \neq 0$). Since $p^k(p^{2k}+1)/2$ is equal to the degree of $T_+(X, 1)$, we conclude that this polynomial splits over $\mathrm{GF}(p^n)$.

Now recall that we found the solutions for $T_+(X, 1) = 0$ in the form $x = \alpha + v^2\alpha^{-1} + 2v = \alpha^{-1}(\alpha + v)^2$ with $\alpha \in \mathrm{GF}(p^n)$. Also, $x + 1 = \alpha + v^2\alpha^{-1} - 2v = \alpha^{-1}(\alpha - v)^2$ since $2v + 1 = -2v$. Thus, both $x$ and $x+1$ are squares in $\mathrm{GF}(p^n)$ if and only if $\alpha$ has this property. Equations (3) result in $\alpha^{(p^n-1)/2} = (-1)^{p^k+1} = 1$ and $\alpha^{(p^n-1)/2} = (-v^2)^{p^k-1} = 1$ which means that $\alpha$ is a square in $\mathrm{GF}(p^n)$.

Now we prove that polynomial $T_-(X, 1)$ does not have any roots in the set of squares of $\mathrm{GF}(p^n)$. Using the substitution $x = \alpha^{-1}(\alpha + v)^2$ exactly the same way as we did for $T_+(X, 1)$, we obtain

$$
\begin{aligned}
T_-(x,1) &= \frac{(\alpha+v)^{2d} - (\alpha-v)^{2d}}{\alpha^d} + \left(\frac{(\alpha+v)^{2d} - (\alpha-v)^{2d}}{\alpha^d}\right)^{p^k} \\
&= 2v\left(\alpha^{(p^{2k}-1)/2} + \alpha^{-(p^{2k}-1)/2} + \alpha^{p^k(p^{2k}-1)/2} + \alpha^{-p^k(p^{2k}-1)/2}\right) \\
&= 2v\left(\alpha^{(p^{2k}-1)/2} + \alpha^{p^k(p^{2k}-1)/2}\right)\left(\alpha^{-(p^{2k}-1)(p^k+1)/2} + 1\right) \quad .
\end{aligned}
$$

Thus, $T_-(x, 1) = 0$ if and only if

$$
\alpha^{(p^{2k}-1)(p^k-1)/2} = -1 \quad \text{or} \quad \alpha^{(p^{2k}-1)(p^k+1)/2} = -1 \quad . \tag{4}
$$

Denote $g = \gcd(p^{8k}-1, (p^{2k}-1)(p^k \pm 1))$ and note that $g$ is equal either to $2(p^{2k}-1)$ or $4(p^{2k}-1)$ depending on $p$. Obviously, $\alpha^g = 1$ if the corresponding identity in (4) holds. If $g = 2(p^{2k}-1)$ then $\alpha^{p^{2k}-1} = -1$ (since supposing $\alpha^{p^{2k}-1} = 1$, none of (4) hold) and $\alpha^{(p^n-1)/2} = (-1)^{(p^{2k}+1)/2} = -1$. Thus, both $\alpha$ and $x = \alpha^{-1}(\alpha+v)^2$ are nonsquares in $\alpha \in \mathrm{GF}(p^n)$. If $g = 4(p^{2k}-1)$ then we need to consider the case when $\alpha^{4(p^{2k}-1)} = 1$ and $\alpha^{2(p^{2k}-1)} \neq 1$. Thus, $\alpha^{p^{2k}-1} = \pm\zeta^{(p^{2n}-1)/4}$, where $\zeta$ is a primitive element of $\mathrm{GF}(p^{2n})$ and

$$
\alpha^{(p^{2n}-1)/2} = \left(\zeta^{(p^{2n}-1)/4}\right)^{(p^n+1)(p^{2k}+1)/2} = \left(\zeta^{(p^{2n}-1)/2}\right)^{(p^n+1)(p^{2k}+1)/4} = -1
$$

since $(p^n+1)(p^{2k}+1)/4$ is odd. Therefore, both $\alpha$ and $x = \alpha^{-1}(\alpha+v)^2$ are nonsquares in $\mathrm{GF}(p^{2n})$ so when $x \in \mathrm{GF}(p^n)$ then it is also a nonsquare in $\mathrm{GF}(p^n)$.

Finally, suppose $T_+(x, c_1) = T_+(x, c_2) = 0$ for some $x \in \mathrm{GF}(p^n)$ and $c_1, c_2 \in \mathrm{GF}(p^k)^*$ with $c_1 \neq c_2$. In particular, we proved that $x + c_1$ is a square in $\mathrm{GF}(p^n)$. Without loss of generality, assume $c_2 - c_1 = 1$ and write

$$\begin{aligned} T_+(x, c_1) - T_+(x, c_2) &= (x + c_1)^d - (x + c_1 + 1)^d + (x + c_1)^{dp^k} - (x + c_1 + 1)^{dp^k} \\ &= T_-(x + c_1, 1) = 0 \end{aligned}$$

that requires $x + c_1$ to be a nonsquare in $\mathrm{GF}(p^n)$. This contradiction shows that roots of $T_+(x, c_1)$ and $T_+(x, c_2)$ lie in disjoint subsets of $\mathrm{GF}(p^n)$. □

The following corollary immediately follows from Proposition 3.

**Corollary 1.** *Let $n = 4k$ and take any $c \in \mathrm{GF}(p^k)^*$. Then polynomial $T_+(X^2, c)$ splits over $\mathrm{GF}(p^n)$ and $x^2 + c$ is a square in $\mathrm{GF}(p^n)$ for every $x \in \mathrm{GF}(p^n)$ with $T_+(x^2, c) = 0$. On the other hand, polynomial $T_-(X^2, c)$ does not have any roots in $\mathrm{GF}(p^n)$. Moreover, for different values of $c$, the corresponding roots of $T_+(X^2, c)$ belong to disjoint subsets.*

Note that taking $c = 0$ in polynomial $T_+(X, c)$ results in

$$T_+(x, 0) = 2\left(x^{(p^{2k}+1)/2} + x^{p^k(p^{2k}+1)/2}\right)$$

which is equal to zero if and only if $x^{(p^{2k}+1)(p^k-1)/2} = -1$. The latter equation has $(p^{2k} + 1)(p^k - 1)/2$ solutions in $\mathrm{GF}(p^n)$. On the other hand, for any $b \in \mathrm{GF}(p^n)$ with $\mathrm{Tr}_k^{2k}\left(b^{p^{2k}+1}\right) = 0$ we have $T_+(b^2, 0) = \mathrm{Tr}_k^{2k}\left(b^{p^{2k}+1}\right) = 0$ and the number of $b \neq 0$ with such a property is equal to $(p^{2k} + 1)(p^k - 1)$. Since squaring defines a 2-to-1 mapping on the set of $b \in \mathrm{GF}(p^n)^*$ with $\mathrm{Tr}_k^{2k}\left(b^{p^{2k}+1}\right) = 0$, we conclude that all nonzero roots of $T_+(X, 0)$ are exactly those $b^2$ for all $b \in \mathrm{GF}(p^n)^*$ with $\mathrm{Tr}_k^{2k}\left(b^{p^{2k}+1}\right) = 0$. Thus, all nonzero roots of $T_+(X^2, 0)$ are exactly those $b \in \mathrm{GF}(p^n)^*$ with $\mathrm{Tr}_k^{2k}\left(b^{p^{2k}+1}\right) = 0$ and the total number of roots is $(p^{2k}+1)(p^k-1)$.

Suppose $T_+(x, 0) = T_+(x, c) = 0$ for some $x \in \mathrm{GF}(p^n)$ and $c \in \mathrm{GF}(p^k)^*$. By Proposition 3, $x$ is a square in $\mathrm{GF}(p^n)$. Then

$$T_+(x, 0) - T_+(x, c) = x^d - (x + c)^d + x^{dp^k} - (x + c)^{dp^k} = T_-(x, c) = 0$$

that requires $x$ to be a nonsquare in $\mathrm{GF}(p^n)$. This contradiction shows that there are no common roots of $T_+(x, 0)$ and $T_+(x, c)$ in $\mathrm{GF}(p^n)$.

**Corollary 2.** *Let $n = 4k$ and take any $b \in \mathrm{GF}(p^n)$. Then polynomials $T_+(b^2, Y)$ and $T_-(b^2, Y)$ have a unique root in $\mathrm{GF}(p^k)$. Moreover, $b^2 + y$ is a square in $\mathrm{GF}(p^n)$ for every $y \in \mathrm{GF}(p^k)$ with $T_+(b^2, y) = 0$. On the other hand, $T_-(b^2, y) = 0$ only for $y = 0$.*

*Proof.* We know that for different values of $c \in \mathrm{GF}(p^k)$, the corresponding roots of $T_+(X^2, c)$ belong to disjoint subsets. Then, by Corollary 1, combining together all roots of $T_+(X^2, c)$ for all $c \in \mathrm{GF}(p^k)^*$ we obtain $p^k(p^{2k} + 1)(p^k - 1)$ elements of $\mathrm{GF}(p^n)^*$. Adding to them $(p^{2k} + 1)(p^k - 1) + 1$ roots of $T_+(X^2, 0)$ gives all $p^n$ elements of $\mathrm{GF}(p^n)$. Thus, for any $b \in \mathrm{GF}(p^n)$, polynomial $T_+(b^2, Y)$ has a unique root in $\mathrm{GF}(p^k)$. The rest follows from Corollary 1. □

## 3 New Class of Binomial Bent Functions

In this section, we prove our main result giving a new binomial weakly regular bent function and calculate its Walsh transform coefficients.

**Theorem 1.** *Let $n = 4k$. Then $p$-ary function $f(x)$ mapping* $\mathrm{GF}(p^n)$ *to* $\mathrm{GF}(p)$ *and given by*

$$f(x) = \mathrm{Tr}_n \left( x^{p^{3k}+p^{2k}-p^k+1} + x^2 \right)$$

*is a weakly regular bent function. Moreover, for $b \in \mathrm{GF}(p^n)$ the corresponding Walsh transform coefficient of $f(x)$ is equal to*

$$S_f(b) = -p^{2k} \omega^{\mathrm{Tr}_k(x_0)/4} \ ,$$

*where $x_0$ is a unique solution in $\mathrm{GF}(p^k)$ of the equation*

$$b^{p^{2k}+1} + (b^2+x)^{(p^{2k}+1)/2} + b^{p^k(p^{2k}+1)} + (b^2+x)^{p^k(p^{2k}+1)/2} = 0 \ . \quad (5)$$

*Proof.* Let $\xi$ be a primitive element of $\mathrm{GF}(p^n)$ and also denote $d = p^{3k} + p^{2k} - p^k + 1$. If we let $x = \xi^j y^{p^{2k}+1}$ for $j = 0, \ldots, p^{2k}$ and $y$ running through $\mathrm{GF}(p^n)^*$ then $x$ will run through $\mathrm{GF}(p^n)^*$ in total $p^{2k}+1$ times. Also note that $d - 2 = (p^{2k}-1)(p^k+1)$ and thus, $d(p^{2k}+1) \equiv 2(p^{2k}+1) \pmod{p^n - 1}$. Therefore, the Walsh transform coefficient of the function $f(x)$ evaluated at $b$ is equal to

$$\begin{aligned}
S_f(b) - 1 &= \sum_{x \in \mathrm{GF}(p^n)^*} \omega^{\mathrm{Tr}_n \left( x^{p^{3k}+p^{2k}-p^k+1} + x^2 - bx \right)} \\
&= \frac{1}{p^{2k}+1} \sum_{j=0}^{p^{2k}} \sum_{y \in \mathrm{GF}(p^n)^*} \omega^{\mathrm{Tr}_n \left( \xi^{jd} y^{2(p^{2k}+1)} + \xi^{2j} y^{2(p^{2k}+1)} - b\xi^j y^{p^{2k}+1} \right)} \\
&= \sum_{j=0}^{p^{2k}} \sum_{z \in \mathrm{GF}(p^{2k})^*} \omega^{\mathrm{Tr}_n \left( \left( \xi^{jd} + \xi^{2j} \right) z^2 - b\xi^j z \right)} \\
&= \sum_{j=0}^{p^{2k}} \sum_{z \in \mathrm{GF}(p^{2k})^*} \omega^{\mathrm{Tr}_{2k} \left( \mathrm{Tr}_k^n \left( \xi^{(p^{2k}-1)j} \right) \xi^{(p^{2k}+1)j} z^2 - \left( b\xi^j + b^{p^{2k}} \xi^{jp^{2k}} \right) z \right)} \ ,
\end{aligned}$$

where $z = y^{p^{2k}+1} \in \mathrm{GF}(p^{2k})^*$ is a $p^{2k}+1$-to-1 mapping of $\mathrm{GF}(p^n)^*$ and since

$$\begin{aligned}
&\mathrm{Tr}_{2k}^n \left( \xi^{jd} + \xi^{2j} \right) = \xi^{jd} + \xi^{2j} + \xi^{jdp^{2k}} + \xi^{2jp^{2k}} \\
&= \xi^{(p^{2k}+1)j} \left( \xi^{p^k(p^{2k}-1)j} + \xi^{-(p^{2k}-1)j} + \xi^{-p^k(p^{2k}-1)j} + \xi^{(p^{2k}-1)j} \right) \\
&= \xi^{(p^{2k}+1)j} \mathrm{Tr}_k^n \left( \xi^{(p^{2k}-1)j} \right)
\end{aligned}$$

noting that $dp^{2k} - p^{2k} - 1 \equiv -p^k(p^{2k}-1) \pmod{p^n - 1}$ and also that $\xi^{-(p^{2k}-1)j} = \xi^{p^{2k}(p^{2k}-1)j}$.

Now, by [3, Corollary 3],

$$S_f(b) - 1 = -sp^k \sum_{j=0}^{p^{2k}} \eta\big(\xi^{(p^{2k}+1)j}\big)\omega^{-\mathrm{Tr}_{2k}\left(\frac{\left(b\xi^j + b^{p^{2k}}\xi^{jp^{2k}}\right)^2}{4\mathrm{Tr}_k^n\left(\xi^{(p^{2k}-1)j}\right)\xi^{(p^{2k}+1)j}}\right)} - p^{2k} - 1$$

$$= -sp^k \sum_{j=0}^{p^{2k}} (-1)^j \omega^{-\mathrm{Tr}_k\left(\frac{\mathrm{Tr}_k^n\left(b^{2p^{2k}}\xi^{(p^{2k}-1)j} + b^{p^{2k}+1}\right)}{4\mathrm{Tr}_k^n\left(\xi^{(p^{2k}-1)j}\right)}\right)} - p^{2k} - 1 \ ,$$

where $s = (-1)^k$ if $p \equiv 3 \pmod 4$ and $s = 1$ otherwise, $\eta(\cdot)$ is the quadratic character of $\mathrm{GF}(p^{2k})$ and since $\eta(t) = 1$ for any $t \in \mathrm{GF}(p^k)^*$ and $\mathrm{Tr}_k^n\big(\xi^{(p^{2k}-1)j}\big)$ is nonzero for any $j = 0, \ldots, p^{2k}$ (we will show this soon). We also used the following identities

$$\begin{aligned}
&\mathrm{Tr}_k^{2k}\left(\xi^{-(p^{2k}+1)j}\big(b\xi^j + b^{p^{2k}}\xi^{jp^{2k}}\big)^2\right) = \\
&= \mathrm{Tr}_k^{2k}\left(b^2\xi^{-(p^{2k}-1)j} + b^{2p^{2k}}\xi^{(p^{2k}-1)j} + 2b^{p^{2k}+1}\right) \\
&= \mathrm{Tr}_k^{n}\left(b^{2p^{2k}}\xi^{(p^{2k}-1)j} + b^{p^{2k}+1}\right) \ .
\end{aligned}$$

Using the notations introduced at the beginning of this section, we get

$$S_f(b) = -sp^k\left(\sum_{u\in U_+}\omega^{-\mathrm{Tr}_k(\psi(u))} - \sum_{u\in U_-}\omega^{-\mathrm{Tr}_k(\psi(u))}\right) - p^{2k} \ ,$$

where $\psi(u) = \mathrm{Tr}_k^n\big(b^{2p^{2k}}u + b^{p^{2k}+1}\big)/4\mathrm{Tr}_k^n(u)$. Note that

$$\mathrm{Tr}_k^n(u) = u + u^{p^k} + u^{-1} + u^{-p^k} = u^{-(p^k+1)}\big(u^{p^k+1}+1\big)\big(u + u^{p^k}\big) = 0$$

only if $u^{2(p^k+1)} = 1$ or $u^{2(p^k-1)} = 1$ which leads to $u^2 = 1$ since $\gcd(2(p^k+1), p^{2k}+1) = \gcd(2(p^k-1), p^{2k}+1) = 2$. Thus, $u = \pm 1$ and for these values of $u$ we have $\mathrm{Tr}_k^n(u) = \pm 4 \neq 0$. We conclude that $\mathrm{Tr}_k^n(u) \neq 0$ for any $u \in U$.

Now we ask ourselves how distinct are the sets of values of $\psi(u)$ when $u \in U_+$ and $u \in U_-$. Consider equation $4\psi(u) = c$ for some $c \in \mathrm{GF}(p^k)$ and the unknown $u \in U$. We can rewrite it as

$$\mathrm{Tr}_k^n\left((b^{2p^{2k}} - c)u\right) = -\mathrm{Tr}_k^n\left(b^{p^{2k}+1}\right) \ . \tag{6}$$

First, if $\mathrm{Tr}_k^n\big(b^{p^{2k}+1}\big) = 0$ then equation (6) has the same number of solutions in $U_+$ and $U_-$ for any $c \in \mathrm{GF}(p^k)$ since $-U_+ = U_-$ which means that $S_f(b) = -p^{2k}$ in this case.

From now on assume $\mathrm{Tr}_k^n\big(b^{p^{2k}+1}\big) \neq 0$ that is equivalent to $\mathrm{Tr}_k^{2k}\big(b^{p^{2k}+1}\big) \neq 0$ since $\mathrm{Tr}_k^n\big(b^{p^{2k}+1}\big) = 2\mathrm{Tr}_k^{2k}\big(b^{p^{2k}+1}\big)$. By Proposition 2, equation (6) has the same number of solutions in $U_+$ and $U_-$ except when $\mathrm{Tr}_k^n\big((b^{2p^{2k}} - c)^{(p^{2k}+1)/2}\big) \neq 0$ and $\mathrm{Tr}_k^n\big(b^{p^{2k}+1}\big) = \pm\mathrm{Tr}_k^n\big((b^{2p^{2k}} - c)^{(p^{2k}+1)/2}\big)$. It is obvious that $c = 0$ gives a solution. Further,

$$\begin{aligned}\mathrm{Tr}_k^n\left((b^{2p^{2k}} - c)^{(p^{2k}+1)/2}\right) &= \mathrm{Tr}_k^n\left((b^2 - c)^{(p^{2k}+1)/2}\right)\\ &= \mathrm{Tr}_k^{2k}\left((b^2 - c)^{(p^{2k}+1)/2} + (b^2 - c)^{(p^n-1)/2+(p^{2k}+1)/2}\right)\end{aligned}$$

is equal to zero if $(b^2-c)^{(p^n-1)/2} = -1$. Therefore, we are interested in the values of $c \in \mathrm{GF}(p^k)$ with

$$(b^2 - c)^{(p^n-1)/2} = 1 \quad \text{and} \quad \mathrm{Tr}_k^{2k}\left(b^{p^{2k}+1} \pm (b^2 - c)^{(p^{2k}+1)/2}\right) = 0\ .$$

By Corollary 2, $\mathrm{Tr}_k^{2k}\big(b^{p^{2k}+1} - (b^2 - c)^{(p^{2k}+1)/2}\big) = T_-(b^2, -c) = 0$ only for $c = 0$ and $\mathrm{Tr}_k^{2k}\big(b^{p^{2k}+1} + (b^2 - c)^{(p^{2k}+1)/2}\big) = T_+(b^2, -c) = 0$ for a unique value $c \in \mathrm{GF}(p^k)$ (for this root, $b^2 - c$ is a square in $\mathrm{GF}(p^n)$, i.e., $(b^2 - c)^{(p^n-1)/2} = 1$). Thus, by Proposition 2,

$$S_f(b) = -sp^k\left(-sp^k + sp^k\omega^{\mathrm{Tr}_k(x_0)/4}\right) - p^{2k} = -p^{2k}\omega^{\mathrm{Tr}_k(x_0)/4}\ ,$$

where $x_0$ is a unique solution in $\mathrm{GF}(p^k)$ of $T_+(b^2, x) = 0$. □

In some cases, we can find exact solution of (5). Assume $b^2 \in \mathrm{GF}(p^{2k})$. Then $b^{2(p^{2k}+1)} = b^4$ giving $b^{p^{2k}+1} = \pm b^2$. It was shown that $x = 0$ is the only solution of (5) if and only if $\mathrm{Tr}_k^{2k}\big(b^{p^{2k}+1}\big) = 0$ that in our case is equivalent to $\mathrm{Tr}_k^{2k}\big(b^2\big) = 0$. In general, (5) takes on the form

$$\begin{aligned}&\pm\left(b^2 + b^{2p^k}\right) + (b^2 + x)^{(p^{2k}-1)/2}\left(b^2 + x + b^{2p^k} + x^{p^k}\right) = 0 \quad \text{or}\\ &\left((b^2 + x)^{(p^{2k}-1)/2} \pm 1\right)\left(b^2 + b^{2p^k}\right) + 2x(b^2 + x)^{(p^{2k}-1)/2} = 0\end{aligned}$$

and is satisfied by exactly one nonzero $x \in \mathrm{GF}(p^k)$ which is $x = -\mathrm{Tr}_k^{2k}\big(b^2\big)$. Therefore, if $b^2 \in \mathrm{GF}(p^{2k})$ then $T_+(b^2, x_0) = 0$ for $x_0 = -\mathrm{Tr}_k^{2k}\big(b^2\big)$.

## References

1. Kumar, P.V., Scholtz, R.A., Welch, L.R.: Generalized bent functions and their properties. Journal of Combinatorial Theory, Series A **40**(1) (September 1985) 90–107
2. Hou, X.D.: $p$-Ary and $q$-ary versions of certain results about bent functions and resilient functions. Finite Fields and Their Applications **10**(4) (October 2004) 566–582
3. Helleseth, T., Kholosha, A.: Monomial and quadratic bent functions over the finite fields of odd characteristic. IEEE Trans. Inf. Theory **52**(5) (May 2006) 2018–2032
4. Niho, Y.: Multi-Valued Cross-Correlation Functions Between Two Maximal Linear Recursive Sequences. PhD thesis, University of Southern California, Los Angeles (1972)

5. Helleseth, T.: A note on the cross-correlation function between two binary maximal length linear sequences. Discrete Mathematics **23**(3) (1978) 301–307
6. Helleseth, T., Rong, C., Sandberg, D.: New families of almost perfect nonlinear power mappings. IEEE Trans. Inf. Theory **45**(2) (March 1999) 475–485